\documentclass[onecolumn,showpacs,amsmath,amssymb,prd,nofootinbib]{revtex4-2}
\usepackage{graphicx}
\usepackage{epsfig}
\usepackage{enumerate}

\def\beq{\begin{equation}}
\def\eeqno#1{\label{#1}\end{equation}}

\def\rarrow{\rightarrow }

\def\dleft{\rlap{{\it D}}\raise 8pt
\hbox{$\scriptscriptstyle\Leftarrow$}}
\def\dright{\rlap{{\it
D}}\raise 8pt\hbox{$\scriptscriptstyle\Rightarrow$}}

\def\az{a_{0}}
\def\azs{a_{0}^2}

\def\l0{\ell_{0}}

\def\rar{\rightarrow}
\def\s{\sigma}

\def\a{\alpha}
\def\b{\beta}
\def\c{\gamma}
\def\l{\lambda}

\def\f{\phi}
\def\t{\theta}

\def\k{\kappa}
\def\r{\rho}
\def\m{\mu}
\def\n{\nu}
\def\z{\zeta}

\def\bo{\bar\omega}

\def\L{\mathcal{L}}
\def\O{\mathcal{O}}
\def\Z{\mathcal{Z}}

\def\R{\mathcal{R}}

\def\d{\delta}

\def\a{\alpha}
\def\xlimin{{x\rarrow\infty \atop{\raise 1pt\hbox to 30pt
{\rightarrowfill}}}}
\def\limlim#1#2{{#1\rarrow #2 \atop{\raise 1pt\hbox to 30pt
{\rightarrowfill}}}}
\def\eps{\epsilon}
\def\vr{{\bf r}}

\def\grad{\vec\nabla}

\def\fs{\f\^*}
\def\gfss{(\gf\^*)^2}

\def\L{\mathcal{L}}

\def\Y{\mathcal{Y}}
\def\G{\mathcal{G}}
\def\T{\mathcal{T}}

\def\gh{g^{1/2}}
\def\M{\mathcal{M}}
\def\tM{\tilde\M}
\def\bM{\bar\M}
\def\m{\mu}
\def\a{\alpha}
\def\b{\beta}
\def\C{\Gamma}
\def\hC{\hat\Gamma}

\def\n{\nu}
\def\Up{\Upsilon}
\def\ten#1#2{\^{#1}\_{#2}}

\def\emn{\eta_{\m\n}}
\def\fh{\hat\f}

\def\_#1{_{\scriptscriptstyle #1}}
\def\^#1{^{\scriptscriptstyle #1}}
\def\baz{\bar a_0}

\def\emn{\eta\_{\m\n}}

\def\gmn{g\_{\m\n}}
\def\Gmn{g\^{\m \n}}

\def\Gma{g\^{\m \a}}
\def\rg{\tilde g}

\def\rgmn{\rg\_{\m\n}}
\def\rGmn{\rg\^{\m\n}}

\def\Gab{g\^{\alpha\beta}}

\def\hGmn{\hat g\^{\m\n}}
\def\hGab{\hat g\^{\a\b}}
\def\hgmn{\hat g\_{\m\n}}

\def\hmn{h\_{\m\n}}
\def\bg{\bar g}

\def\bgmn{{\bar g}\_{\m\n}}
\def\bGmn{{\bar g}\^{\m\n}}
\def\der#1{\_{,#1}}

\def\lM{\ell\_M}

\def\oot{\frac{1}{2}}

\def\grp{\breve{g}}
\def\drp{\breve{\delta}}

\def \hij {h\_{ij}}
\def\hhij{\hat h\_{ij}}
\def\hsij{h\^*\_{ij}}
\def\hs{h\^*}

\def\epg{8\pi G}
\def\spg{16\pi G}

\def\gfss{(\nabla\fs)\^2}
\def\vrf{\varphi}
\def\vrfzs{\vrf\der{0}\^2}
\def\gvrfs{(\grad\varphi)^2}
\def\CC{\mathcal{C}}

\def\MG{\M\_G}
\def\sC{{^*\C}}
\def\shC{{^*\hat\C}}
\def\sg{\textsl{g}}
\def\sg{\mathfrak{g}}
\def\LCDM{$\Lambda{\rm CDM~}$}
\def\VGM{VGMOND}
\begin{document}
\title{Bimetric MOND as a framework for variable-$G$ theories: Local systems and cosmology}

\author{Mordehai Milgrom}
\affiliation{Department of Particle Physics and Astrophysics, Weizmann Institute}

\begin{abstract}
Bimetric MOND (BIMOND) is used as a platform for variable-$G$ theories that have MOND-specific idiosyncrasies. E.g.,
MOND premises dictate return to standard dynamics in the high-acceleration limit, predicting the standard value of $G$ for high-acceleration systems. This automatically ensures compliance of such theories with all the constraints on inconstancy of $G$ that emerge from the study of high-acceleration systems: geophysics, solar system, pulsars, supernovae, stellar evolution, emission of gravitational waves, etc. In MOND, constraints deduced from such phenomena have no bearing on possible $G$ variability in cosmology.
My guiding motivation is to see if such theories may account for some roles of dark matter in cosmology; e.g., in accounting for the expansion history of the Universe in the matter-dominated era, by having a $G_e\approx 2\pi G$ govern the later stages of the expansion, instead of invoking matter density $\approx 2\pi\times$ baryon density.
BIMOND is a relativistic MOND theory whose gravitational degrees of freedom are two metrics: $\gmn$, to which matter couples minimally, and $\hgmn$, defining the geometry of a twin sector, presumably with its own ``twin-matter.''
The BIMOND gravitational action for the metric-symmetric subclass is $I\^B\propto G^{-1}\int d^4x ~[|g|\^{1/2}R+|\hat g|\^{1/2}\hat R+v(g,\hat g)\lM\^{-2}\tM]$, where the metric-interaction term, $\tM$, is a function of dimensionless scalars constructed from $\lM C\ten{\a}{\b\c}$, where the ``relative-acceleration'' tensor, $C\ten{\a}{\b\c}\equiv \C\ten{\a}{\b\c}-\hat\C\ten{\a}{\b\c}$, is the difference of the Levi-Civita connections, and $\lM=c^2/\az$ is the MOND length. Locally, where the random fluctuations in the two metrics are different, $\tM$ encapsulates MOND effects in galactic systems, and the average of $\tM$ on cosmological scales acts as a (dynamical) dark energy.
Without adding degrees of freedom, or new dimensionful constants, BIMOND can be extended to a class of theories that entail what is best described as phenomenon-dependence of Newton's constant, $G$: The Lagrangian density in $I\^B$ is multiplied by $1+\MG$, where $\MG$ too is a function of the above scalars. $G_e\equiv G/(1+\MG)$ plays the role of the effective gravitational constant.
I consider the standard and the Einstein-Palatini formulations for the above action.
My aim here is to present the general framework, while specific predictions would depend crucially on details of the structure of the theory -- e.g., on the choice of variables, and the dependence of $\tM$ and $\MG$ on them -- and, in cosmology, on unknown inputs: cosmological initial conditions, material content of the twin sector, etc. I cannot yet present a consistent model that complies with all the observations in cosmology, including the expansion history, with all its details.
Instead, I describe some examples of theories in the class that predict different values of $G_e$ in different circumstances, including one where $G$ takes its standard value for all subcosmological systems -- even if they are deep in the MOND regime. I also discuss scenarios in which $G_e\approx G$ in the early Universe, as required by constraints from big bang nucleosynthesis, but with $G_e> G$ setting in at later times, where it can affect the expansion history during the matter-dominated era.

\end{abstract}
\maketitle

\section{Introduction   \label{introduction}}
MOND \cite{milgrom83} is a paradigm that accounts for the mass discrepancies in galactic systems without dark matter. Instead, it posits departure from standard dynamics -- Newtonian dynamics and general relativity (GR) -- at low accelerations, at and below the MOND acceleration constant$, \az$. In the ``deep-MOND''  limit (DML), i.e., much below $\az$, MOND dictates that dynamics of systems governed by gravity become spacetime scale invariant -- i.e., invariant under $(\vr,t)\rar\l(\vr,t)$ -- at least in the nonrelativistic (NR) limit \cite{milgrom09}. Another tenet of MOND is that at high accelerations, much above $\az$, dynamics returns rapidly to standard dynamics.
This means, formally, that a relation deduced with MOND, which typically contains $\az$, should tend rapidly, as $\az\rar 0$, to the standard relation describing the same phenomenon. This limit corresponds to a situation where the system attributes with the dimensions of acceleration are much larger than $\az$.
Reviews of MOND, with different emphases and extent, can be found in Refs. \cite{fm12,milgrom14,milgrom20a,mcgaugh20,merritt20,bz22,fd25}.
\par
The value of $\az$ -- as determined from its various appearances in galactic phenomena and laws predicted by MOND -- is close to cosmologically-significant accelerations, such as those defined by the Hubble-Lema$\hat{\rm i}$tre constant, $H_0$, and by the ``cosmological constant'' or ``dark-energy'' curvature, $\Lambda$ \cite{milgrom83,milgrom99,milgrom20}:
\beq \baz\equiv 2\pi\az\approx c^2(\Lambda/3)^{1/2}, ~~~~~{\rm and}~~~~~\baz\approx c H_0.  \eeqno{coinc}
\par
Either of these coincidences may be a crucial pointer to the underpinnings of MOND \cite{milgrom99,milgrom20}. For example, it may indicate that MOND phenomenology in local systems -- such as galactic systems -- cannot be understood, in a fundamental manner, separately from cosmology. If indeed fundamental, this coincidence tells us that either the cosmological state and evolution of the Universe at large strongly enters local dynamics, or that cosmology and local dynamics are affected by the same agent whose characteristic acceleration enters both cosmology and galactic dynamics. Various ways in which this can happen are sketched in Ref. \cite{milgrom20}.
\par
If we were convinced that the coincidence is fundamental -- which is yet to be established -- one would necessarily require from a fundamental MOND theory that it accounts for this coincidence.
\par
But even if it is not fundamental, the mere numerical coincidence has some important implications for how MOND enters relativistic phenomena \cite{milgrom20}.
In particular, one such implication, greatly relevant to our present discussion, and, more generally, to the construction of relativistic-MOND theories, is that there are no systems that are both of strong (relativistic) gravity ($MG/\ell\not \ll c^2$), and of low acceleration ($MG/\ell^2\lesssim \az$), {\it except the Universe at large}. (Here, $M$ is the characteristic mass, and $\ell$ the characteristic size of the system.) This is because such two requirements, with the numerical coincidence (\ref{coinc}) would imply that $\ell>\lM\sim\ell\_U$, where $\lM\equiv c^2/\az$ is the ``MOND length'', and $\ell\_U$ is the characteristic size in cosmology (the Hubble distance, or the de Sitter radius associated with $\Lambda$).
\par
By and large, MOND accounts very well for dynamics in galactic systems -- in fact, it had accurately predicted many of the regularities that were subsequently observed in such systems (see Refs. \cite{fm12,milgrom14,milgrom20a,mcgaugh20,merritt20,bz22,fd25}).
The question then naturally arises whether some relativistic MOND theory can also account for the observations in cosmology, without dark matter.
\par
The exact structure-formation scenario in MOND is still moot. But several works have shown \cite{sanders98,sanders08,mcgaugh15,eppen22,mcgaugh24} that it occurs, generically, in a different way than in the \LCDM paradigm, with galaxies and larger-scale structure forming rather earlier, as, in fact, appears to be vindicated by recent observations \cite{mcgaugh24}.
\par
There are also natural explanations for the appearance of a cosmological constant, or ``dark energy'', within various MOND theories (\cite{milgrom20}, and Sec. \ref{cosmology} below).
\par
Putting aside, then, the important question of what replaces (in MOND) the role of dark matter in CMB fluctuations, and structure formation, we ask, as a limited question, whether MOND can account, not only for the role of dark energy, but also for that of dark matter, in dictating the observed expansion history of the Universe. A possible pointer toward effecting this end is the fact that in this context, the densities of (NR) baryons, $\r_b$, and of DM, $\r\_{DM}$, enter in the combination $G(\r_b+\r\_{DM})$, and, to account for the expansion history with dark matter it is required that $\r\_M=\r_b+\r\_{DM}\approx 2\pi\r_b$  (e.g., Ref. \cite{planck20}). Thus, the same history results, without dark matter, if the effective $G$ value that governs the expansion history during the relevant epoch is $G\_e\approx 2\pi G$.
\par
A further motivation for such attempts comes from the fact -- which is still a mystery -- that the two densities are so near each other, $\r\_{DM}\approx 5\r\_b$, even though these densities were determined, for all we know, by very different physical processes, taking place at very different times, when the physical conditions were very different. Such a coincidence would result naturally if the two are not really two distinct matter components, but if $\r\_M$ is related to $\r_b$ by a theory that implies that their (possibly inconstant) ratio is of order unity.
\par
Seeking an explanation in such a vein would involve a ``variable-$G$'' theory (VGT), constructed on the principles of MOND.
\par
I understand the term VGT to mean the following:
Starting from some relativistic, Lagrangian-based theory of gravity -- usually, GR, but here bimetric MOND (BIMOND) -- with a Lagrangian $\propto G^{-1}\L$, replace $\L$ with $f\L$, where $f$ is a function of some dynamical, gravitational degrees of freedom (DoFs). $f$ is then dimensionless, and normalized such that $f\equiv 1$ in the limit when the original theory is restored.
\par
Predictions of such theories cannot, generally, be described as those of standard dynamics only with a different, spacetime-dependent,\footnote{As always, variability of $G$ is understood as the variability of its value as measured in atomic units.} effective $G_e=G/f$. But, such an ansatz is sometimes justified approximately when testing and constraining the possible inconstancy of $G$: One does, usually, consider some phenomenon -- e.g., big bang nucleosynthesis, stellar evolution, pulsar timing, etc. -- and analyzes it under the above ansatz. A comprehensive review of such tests can be found in Ref. \cite{uzan25} (Sec. 6 there is dedicated to tests of, and limits on, possible inconstancy of $G$).
\par
Since MOND introduces a new dimensionful ``dividing constant'',\footnote{By a ``dividing constant,'' I mean a physical constant, $\CC$, that, among other roles, defines a dividing line: Systems whose attributes, $\Y$, with the dimensions of $\CC$, behave differently -- e.g., obey different scaling relations -- depending on whether $\Y/\CC$ is small or large. Usually, $\CC$ itself appears ubiquitously in relations that hold in one of the limits, and disappears in those valid in the other limit. This clearly include $c$ in relativity, $\hbar$ in quantum physics, and $\az$ in MOND, but not $G$, or Boltzmann's constant, which are only ``conversion'' constants, and do not play such a role.} $\az$, together with $c$ they define a fourfold classification of phenomena and systems, according to whether they are relativistic or not, and according to whether they are high-acceleration or not, in the MOND context. All subcosmological phenomena for which tight constraints on $G$ inconstancy exist, are either NR, or characterized by very-high-accelerations, because the ``coincidence'' (\ref{coinc}) tells us that relativistic, subcosmological systems are perforce of high accelerations. This puts all of them in separate classes from cosmology. This makes it natural in MOND to have $G$-inconstancy effects in cosmology, without them showing up in subcosmological systems.\footnote{When dealing with a VGT that modifies GR (and retains the standard Newtonian limit), but does not introduce new dimensionful constants -- as is the case in the iconic Jordan-Brans-Dicke VGT \cite{brans14} -- we have a twofold classification of systems according to whether they are relativistic or not. The strength of $G$-inconstancy effects then depends on how relativistic the system is, which puts cosmology in the same class as black holes, pulsar phenomena, compact-stars mergers with the emission of gravitational waves, etc.
Then, tight constraints obtained from relativistic, subcosmological systems and phenomena, bear, directly, on $G$ inconstancy in cosmology.}
Such an isolation of cosmology as the only phenomenon in the strong-gravity-low-acceleration quadrant would underlie any VGT theory based on MOND. Here I shall demonstrate it specifically with BIMOND-based VGTs.
Such theories may have additional idiosyncrasies that bear on the issue, and that are not necessarily shared by other MOND-based VGTs.
\par
A relativistic theory whose NR limit reproduces MOND phenomenology, and that goes a long way in accounting for cosmology, is described in Ref. \cite{sz21}. This theory harks back to Bekenstein's TeVeS theory \cite{bekenstein04}, employing the same gravitational DoFs, but employing a different Lagrangian. In particular, dark matter in GR cosmology is replaced, in this theory, by a k-essence mechanism \cite{scherrer04}, where the scalar field that appears anyhow in TeVeS-rooted theories, plays the role of the k-essence field. Reference \cite{blanchet24} proposes another theory that reproduces both MOND phenomenology and cosmology, based on a khronon field as the k-essence scalar.
In Ref. \cite{milgrom22}, I briefly outlined a possible way to achieve a similar effect with bimetric MOND (BIMOND).
\par
Here I take a different approach that still starts from BIMOND, as an example.
BIMOND \cite{milgrom09b,milgrom22} is a class of effective, relativistic, modified-gravity, MOND theories. They employ two metrics as gravitational DoFs, with matter in the sector we are part of coupling minimally to one metric, $\gmn$. The other  metric, $\hgmn$, possibly comes with its own ``twin matter'' sector. Such a putative twin matter couples to $\hgmn$, but not directly to $\gmn$ or to matter. The two metrics couple directly to each other via a function, $\tM$, of the ``relative acceleration scalars'' constructed from the difference in the affine connections of the two metrics.
A more detailed description of BIMOND is given in Sec. \ref{recap}.
\par
It has been shown in Refs. \cite{milgrom09b,milgrom22} that a large subclass of BIMOND theories have a NR limit that yield MOND phenomenology as observed, including ``correct'' gravitational lensing by galactic systems.
\par
Cosmology in BIMOND has been studied only superficially (e.g., in Refs. \cite{milgrom09b,milgrom10b,clifton10}).
An important point that we can make, generally, is that BIMOND theories account naturally for the appearance of ``dark energy'' in cosmology, whose contribution to the curvature is of the order of $\lM^{-2}$, accounting for the first near equality in Eq. (\ref{coinc}) (see Sec. \ref{cosmology}).
\par
Here, I propound a generalization of BIMOND, where I keep $\az$ as the sole dimensionful constant besides $c$ and $G$, and the two metrics as the only DoFs that mediate gravity to matter and twin matter. I also build on the same ``relative-acceleration scalars'' through which these metrics interact in BIMOND. I then modify the BIMOND action, by replacing $G$ in the Lagrangian by $G/(1+\MG)$, where $\MG$ is another function of these scalars. This results in a larger family of theories -- \VGM~ -- which constitute VGTs in the sense defined above.
\par
Strictly speaking, these \VGM~ theories introduce spacetime dependence of $G$ via that of $\MG$. But, in light of what was said above, in many instances it may be more useful to think of them as introducing phenomenon- or system-dependent $G$.
\par
{\it I do not demonstrate here that there are theories in this class that conform to all constraints from cosmology.} At present, the suggested class of theories remains an interesting framework for VGTs that aligns with MOND, from which they borrow heavily.
\par
Our general observations above concerning MOND theories, apply, of course, to the BIMOND-based VGTs I shall discuss here. To recapitulate this important point:
In these theories, the return to standard gravity at high accelerations -- a tenet of MOND -- ensures compliance with the known constraints on $G$ inconstancy from subcosmological phenomena (see Ref. \cite{uzan25}). These include, Earth-bound phenomena, lunar ranging, solar-system dynamics, different aspects of pulsar timing, stellar structure and evolution, gravitational-wave radiation from coalescing compact objects, etc. And yet, in cosmology these theories can exhibit effective $G$ inconstancy.
\par
In BIMOND, specifically, there is an additional idiosyncrasy that affords an interesting leverage: It is that in situation of equality of the two metrics, all the scalar variable vanish, and, more generally, the departure from equality is a parameter that can control the departure of $G_e$ from $G$. I use this fact to suggest how in a cosmology that starts with an exactly metric-symmetric
initial conditions, initially, $G_e=G$, which would comply with the constraints from big bang nucleosynthesis, with a departure taking place only at later epochs, when asymmetric random fluctuations develop and increase.
\par
Note, in this connection, that the accounts of galaxy formation in Refs. \cite{sanders98,sanders08,mcgaugh15,eppen22,mcgaugh24}, and others, are not based on some relativistic theory that takes into account the cosmological expansion, together with the gravitational collapse of structure.
In particular, they all assume that the acceleration that determines the collapse dynamics is only that contributed by the local overdensity, and ignore the background acceleration associated with the expansion. Metric-symmetric versions of BIMOND justify this ansatz formally. What enter their MOND aspects that govern the dynamics of {\it local} systems are the ``relative acceleration'' scalars, in which the background expansion cancels out (see Sec. \ref{cosmology}).
\par
As regards low-acceleration, local systems, such as virialized galactic systems, I will require that BIMOND itself (without the VGT extension) accounts for galactic dynamics, and I shall, indeed, describe versions of the proposed \VGM~ in which the variable-$G$ aspects are not expressed in NR systems, which include all galactic systems.\footnote{It has been proposed that MOND phenomenology in galactic systems results from dependence of $G$ on acceleration (e.g., Refs. \cite{kazanas18,kazanas23}). I emphasize that this is not what I am suggesting here.}
\par
I end this introduction by a general comment on relativistic MOND theories. Inasmuch as they tend quickly to GR for high-acceleration systems, such theories are not required to account for relativistic objects whose dynamics depart from GR, except for cosmology, which is a unique system.
I think then that we may be more forgiving to such theories  of some issues that might only appear in applications to local systems (such as Ostrogradky's instabilities, or the appearance of ghosts (see Ref. \cite{dambrosio20} for a discussion of such issues with the first version of BIMOND). For such systems, the theory practically coincides with GR, which is free of these issues. And if such theories turn out not to be ``fundamental'' -- which known working theory is? -- they can at least serve as useful heuristic tools. In the present discussion, they can limelight the idiosyncracies of MOND based VGTs.
\par
In Sec. \ref{recap}, I recap, with some new insights, some relevant BIMOND results rom Refs. \cite{milgrom09b,milgrom22}, including the emergence in BIMOND of ``dark energy'' of the right ballpark magnitude, and the constraints on the choice of interaction scalar variables from gravitational lensing. Section \ref{framework} presents the extension of BIMOND to the \VGM~ class of theories. These theories are interpreted in this section with the standard ``metric'' approach in which the two metrics are the only gravitational DoFs. In Sec. \ref{EP} the same action is discussed in the Einstein-Palatini framework, where the metrics and the connections from which the curvature tensors are constructed are treated as independent DoFs.
Section \ref{constraints} discusses some general implications, and some specific examples.
In Sec. \ref{cosmology}, I describe some preliminary thoughts on cosmology.
In Sec. \ref{discussion}, I list several remaining open questions and issues concerning this class of \VGM~ theories.

\section{BIMOND recap  \label{recap}}
Here I give a brief recap of the basic elements of BIMOND, as discussed in more detail in Refs. \cite{milgrom09b,milgrom22}.
(The nomenclature and exact choice of details of the action have changed somewhat from treatment to treatment.) The basic BIMOND gravitational action for the subclass of theories that are symmetric in the two sectors can be taken as ({\it putting $c=1$ in what follows})
\beq I\^B=-\frac{1}{\spg}\int d^4x ~[|g|\^{1/2}R+|\hat g|\^{1/2}\hat R+v(g,\hat g)\lM^{-2}\tM(\Z_1,\Z_2,...)],  \eeqno{muba}
where the first two terms are the standard Einstein-Hilbert actions for the two sectors, while the third term encapsulates the interaction between the metrics. The dimensionless variables $\Z_m=\lM^2 S_m$, where $S_m$ are scalars quadratic in the ``relative-acceleration tensors''
 \beq C\ten{\a}{\b\c}\equiv \C\ten{\a}{\b\c}-\hat\C\ten{\a}{\b\c},   \eeqno{narate}
 i.e., the differences of the Levi-Civita connections of the two metrics. The scalars are of the form
\beq S=Q\ten{\b\c\m\n}{\a\l}C\ten{\a}{\b\c}C\ten{\l}{\m\n},\eeqno{zababa}
where the contraction coefficients $Q\ten{\b\c\m\n}{\a\l}$ are constructed in a symmetric way from the two metrics (see more details in Sec. \ref{scalars}).
To avoid coupling of the two metrics in the volume element itself, I work with the volume element of the interaction term
\beq  v(g,\hat g)=|g|\^{1/2}+|\hat g|\^{1/2},   \eeqno{volel}
for which the two sectors decouple for interesting cases where $\tM$ is constant.
The matter actions, not shown here, are the standard ones with matter in each sector coupling minimally to its own metric.\footnote{More general versions may be, and have been, considered; e.g., with lack of symmetry between the two sectors, with different volume elements, with nonquadratic scalars, etc.}

\par
Choosing the proper form of $\tM$ yields a NR limit that satisfies the MOND tenets, and accounts for dynamics in galactic systems, without dark matter. Furthermore, for a large class of scalar-variable choices (see Sec. \ref{scalars} below), ``correct'' gravitational lensing is predicted \cite{milgrom09b,milgrom22}.
\par
The ``correspondence principle'' whereby the theory goes to GR in the limit of high acceleration -- formally by taking $\Z\gg 1$ -- is ensured by having $\tM$ becoming a (dimensionless) constant, $\tM\rar\tM_\infty$, in this limit. Then the two sectors decouple [with the choice (\ref{volel}) of the volume element] and each is governed by GR gravity, with a cosmological constant $\Lambda= -\lM^{-2}\tM_\infty/2$.

\par
If the dimensionless $\tM_\infty$ is of order unity, the appearance of such a cosmological constant in the dynamics of subcosmological, high-acceleration systems can at most play a very minor role.
Its contribution to the local accelerations would be of order $\az(\ell/\lM)\ll\az$, where $\ell$ is the system's size.
\par
Another interesting limit corresponds to exact symmetry between the two sectors.
In this case, $\tM=\tM_0\equiv\tM(0,0,...)$ and, again, we get GR behavior, with a cosmological constant  $\Lambda= -\lM^{-2}\tM_0/2$.
\par
The BIMOND field equations are of the form
\beq G\^{\m\n}+\T\^{\m\n}+\epg T\^{\m\n}\_{M}=0,   \eeqno{melte}
\beq \hat G\^{\m\n}+\hat \T\^{\m\n}+\epg \hat T\^{\m\n}\_{M}=0,   \eeqno{melteha}
where
$G\^{\m\n}$ and $\hat G\^{\m\n}$ are the Einstein tensors of the two metrics, $T\^{\m\n}\_M$ and $\hat T\^{\m\n}\_M$ are the matter and twin-matter energy-momentum tensors, and $\T\^{\m\n}$ and $\hat \T\^{\m\n}$ are those gotten from varying the BIMOND interaction term.
\par
The Bianchi identities, and the related Cauchy problem for BIMOND, are discussed in Ref. \cite{milgrom09b}: While $G\^{\m\n}$ and $\hat G\^{\m\n}$ are identically divergenceless (each with its own covariant divergence), $\T\^{\m\n}$ and $\hat \T\^{\m\n}$ are not identically divergenceless. There are, however, four combined Bianchi identities following from the covariance of the interaction action. They are
\beq (g/\hat g)\^{1/2}{\T\^{\m\n}}\_{;\n}+{{\hat\T}\^{\m\n}}\_{~~~:\n}=0.   \eeqno{bianch}
Here, ``;''  denotes the covariant derivative with respect to $\gmn$, and ``:''  that with respect to $\hgmn$.
\par
Details of how $\T\^{\m\n}$ and $\hat \T\^{\m\n}$ are given in terms of the dependence of $\tM$ on the quadratic scalars are given in Ref. \cite{milgrom22}.
\par
Another class of bimetric extensions of GR that were proposed to account for MOND phenomenology are described in Ref. \cite{bernard15}.

\subsection{The BIMOND scalar variables  \label{scalars}}
Here I describe the scalar variables, their behavior in various relevant circumstances, and constraints on their choice that we already have from observed phenomenology. These are important to appreciate when constructing \VGM~ theories.
\par
Although, in principle, one can employ scalars that are higher order in the $C\ten{\l}{\m\n}$ (they have to be of even order since the indices have to be contracted in pairs), I always concentrate on theories that use only quadratic scalars, which then have the form (\ref{zababa}) [$S=Q\ten{\b\c\m\n}{\a\l}C\ten{\a}{\b\c}C\ten{\l}{\m\n}$].
The contraction coefficients, $Q\ten{\b\c\m\n}{\a\l}$, are constructed from $\gmn$, $\hat g\_{\m\n}$, and their inverses; and from $\d\ten{\a}{\b}$.
\par
There are only five archetypal, independent, basic forms of  $Q\ten{\b\c\m\n}{\a\l}$, one choice of which can be written formally as
\beq Q\ten{\b\c\m\n}{\a\l}=\grp\^{\b\m}\drp\ten{\n}{\a}\drp\ten{\c}{\l},~~~ \grp\^{\b\c}\drp\ten{\m}{\a}\drp\ten{\n}{\l},~~~
\grp\_{\a\l}\grp\^{\b\c}\grp\^{\m\n},~~~\grp\^{\b\m}\drp\ten{\c}{\a}\drp\ten{\n}{\l},~~~
\grp\_{\a\l}\grp\^{\b\m}\grp\^{\c\n},  \eeqno{meutaz}
where $\grp$ is either $g$ or $\hat g$, and $\drp\ten{\a}{\b}$ stands for $\d\ten{\a}{\b}$, $q\ten{\a}{\b}\equiv g\^{\a\s}\hat g\_{\s\b}$,  or its inverse $\hat g\^{\a\s} g\_{\s\b}$.
The general $Q\ten{\b\c\m\n}{\a\l}$ is a linear combination of such tensors, with coefficients that can depend on the scalars $\k\equiv(g/\hat g)\^{1/4}$ and $\bo\equiv \Gmn\hgmn$. (Derivatives of the scalars $\k$ and $\bo$ are expressible in terms of $C\ten{\a}{\b\c}$.)
\par
We shall be interested, in what follows, in situations where the two metrics differ only slightly from a single reference metric, $\bgmn$. Then, to the lowest order in these differences, only $\bgmn$ enters all the contractions. This applies, e.g., to the weak-field, and NR limits on a background of double Minkowski, or, in cosmology, where $\bgmn$ will be a FLRW background metric.
Under such circumstances, there are, strictly, only five independent, quadratic scalars, which can be chosen as
\beq S_1\equiv\bGmn C\ten{\c}{\m\l}C\ten{\l}{\n\c},~~~S_2\equiv\bar C\^{\c}C\_{\c},~~~
 S_3\equiv\bgmn\bar C\^{\m}\bar C\^{\n},~~~ S_4\equiv\bGmn C\_{\m} C\_{\n},
 ~~~S_5\equiv \bg\_{\a\l}\bg\^{\b\m} \bg\^{\c\n}C\ten{\a}{\b\c}C\ten{\l}{\m\n},
 \eeqno{meqaz}
  where
$\bar C\^{\c}\equiv \bGmn C\ten{\c}{\m\n}$ and $C\_{\c}\equiv C\ten{\a}{\c\a}$  are the two traces of $C\ten{\l}{\m\n}$.

\subsubsection{Lensing constraints on the scalars \label{lens}}
As discussed in Ref. \cite{milgrom22}, the requirement of ``correct'' gravitational lensing limits the choice of scalar arguments that we can employ. By ``correct'' lensing, I mean that the NR metric $\gmn$ can be brought (by an appropriate choice of gauge) to the form it has in GR, $g\_{\m\n}=\emn-2\f\d\_{\m\n}$; so that there is only one potential that governs the geodesics of both slow particles and photons.
This seems to hold in nature, at least approximately, when comparing the gravitational potentials deduced for a given body (e.g., a galaxy or a galaxy cluster) from lensing, and from slow-particle dynamics.
\par
Expanding the two metrics around a common Minkowski background, we write to first order in the departures (the weak-field limit)
 \beq g\_{\m\n}=\emn-2\f\d\_{\m\n} +h\_{\m\n},~~~~
 \hat g\_{\m\n}=\emn-2\fh\d\_{\m\n}+\hat h\_{\m\n}.
\eeqno{rutza}
Defining $\f\equiv (\eta\_{00}-g\_{00})/2$ and  $\fh\equiv (\eta\_{00}-\hat g\_{00})/2$, we have $h\_{00}=\hat h\_{00}=0$.
Also, the differences and sums of the potentials
 \beq \fs=\f-\fh, ~~~\hs\_{\m\n}=h\_{\m\n}-\hat h\_{\m\n}~~~\f\^+=\f+\fh, ~~~h\^+\_{\m\n}=h\_{\m\n}+\hat h\_{\m\n}.
 \eeqno{metza}
In the NR limit, we deal with static sources, and the energy-momentum tensors of matter are thus $T\^M\_{\m\n}=\r\d\_{\m 0}\d\_{\n 0}$, and similarly for the hatted one.
\par
The BIMOND field equations imply that the mixed space-time potentials vanish (also follows from the time-reversal symmetry of the static sources). The sum and difference field equations involve, respectively, the sum and difference potentials, which thus decouple. The former is equivalent to the GR equations, with all the gauge freedom. So, as in GR, we can work in a gauge where $h\^+\_{\m\n}=0$. {\it To ensure correct lensing, it is thus necessary and sufficient that the NR field equations imply that $\hsij=0$.}
\par
By examining the field equation that results from variation of the NR action with respect to $\hij$ and $\hhij$, is was shown in Ref. \cite{milgrom22} that a sufficient condition for this is that the scalar arguments of $\tM$ do not contain mixed $\fs-\hsij$ terms.
\par
The NR expressions of the independent scalars in Eq. (\ref{meqaz}) are (ignoring the mixed space-time components of the potentials):
\beq \bar S_1=-2\gfss+\frac{1}{4}\hs\_{kj,i}(2\hs\_{ki,j}-\hs\_{kj,i})-2\fs\der{i}(\hs\_{ik,k}-\oot \hs\der{i}),\eeqno{s1}
\beq \bar S_2=\oot \hs\der{i}(\hs\_{ik,k}-\oot \hs\der{i})-2\fs\der{i}(\hs\_{ik,k}-\oot \hs\der{i}),\eeqno{s2}
\beq \bar S_3=(\hs\_{ik,k}-\oot \hs\der{i})(\hs\_{il,l}-\oot \hs\der{i}),\eeqno{s3}
\beq \bar S_4=4\gfss+\frac{1}{4}\hs\der{i}\hs\der{i}-2\fs\der{i}\hs\der{i},\eeqno{s4}
\beq \bar S_5=10\gfss+\frac{1}{4}\hs\_{kj,i}(3\hs\_{kj,i}-2\hs\_{ki,j})+2\fs\der{i}(\hs\_{ik,k}-\frac{3}{2}\hs\der{i}).\eeqno{s5}
($\hs=\sum\_{k=1}\^3 \hs\_{kk}$ is the spatial trace of $\hsij$).
\par
Because there are two forms of mixed terms, $\fs\der{i}\hs\der{i}$ and $\fs\der{i}\hs\_{ik,k}$, eliminating them leaves us with a three-parameter class of scalars -- call them ``good scalars of the first type'' -- whose general form is \cite{milgrom22}
\beq S_{q,u,v}=qS_q+uS_u+vS_v;~~~~~S_q\equiv\oot(3S_1-2S_2-S_3-S_4+S_5),~~~S_u\equiv\frac{1}{4}(S_1-S_4+S_5),~~~S_v\equiv S_3,   \eeqno{gumsha}
whose NR limit is
\beq \bar S_{q,u,v}=u\gfss+ \frac{u}{8}\hs\_{ij,k}\hs\_{ij,k}+\frac{q}{2}\hs\_{ij,k}\hs\_{ik,j}-v\hs\der{i}\hs\_{ik,k}+ \frac{1}{4}(v-\frac{u}{4})\hs\der{i}\hs\der{i}+(v-\frac{q}{2})\hs\_{ik,k}\hs\_{ij,j}.  \eeqno{bituv}
\par
In Ref. \cite{milgrom22}, I stated, mistakenly, that these are the only scalars compatible with correct lensing; i.e., that it is also necessary to employ only such scalars in the arguments of $\tM$.
This, however is not the case: Reexamining the relevant field equations, I noticed that the condition for ``correct'' lensing, i.e., $\hsij=0$, is satisfied even if $\tM$ depends on scalars, $\tilde S$, that do include mixed, $\fs-\hsij$ terms, but do not include $\gfss$ terms -- provided $\partial\tM(\tilde S,S_1,...)/\partial \tilde S|_{\tilde S=0}=0$. The general form of such scalars is
\beq \tilde S=\a S_2+\b S_4+\c S_5+(2\b+5\c)S_1+\varepsilon S_3; \eeqno{mayure}
call them ``good scalars of the second type''.
\par
There is some overlap between the two types of scalars: those in Eq. (\ref{mayure}), with both $\a=-2\c$ and  $\b=-\c$ are also of the first type, as they contain only terms with $\hsij$. These are also the scalars described by Eq. (\ref{bituv}) with $u=0$.
\par
Since the scalars of the second type vanish everywhere in the NR limit (where $\hsij=0$), even if $\tM$ depends on them with the above proviso, they do not enter NR phenomenology. However, their potential importance in cosmology makes them central in the present \VGM~ context.

\section{BIMOND framework for effective variable-$G$ theories   \label{framework}}
To obtain a variable-$G$ class of theories, we modify the original BIMOND gravitational action (\ref{muba}), taking instead,
\beq I\_G\^B=-\frac{1}{\spg}\int d^4x ~(1+\MG)[|g|\^{1/2}R+|\hat g|\^{1/2}\hat R+(|g|\^{1/2}+|\hat g|\^{1/2})\lM\^{-2}\tM], \eeqno{acGmond}
where $\MG=\MG(\Z\^G_1,\Z\^G_2,...)$, like $\tM$, is a dimensionless function of the variables $\Z\^G_k=\lM^2S\^G_k$, where $S\^G_k$ are ``good'' scalars from among those discussed in Sec. \ref{lens}.
$\MG$ is to encapsulate the inconstancy of the effective gravitational constant $G_e=G/(1+\MG)$.
\par
It is convenient to absorb $(1+\MG)$ into $\tM$, in the third term in the gravitational action, since they depend on the same type of variables; so, define
$\bM\equiv\tM(1+\MG)$, and work with the equivalent form :
\beq I\_G\^B=I\^B(\tM\rar\bM)-\frac{c^4}{\spg}\int d^4x ~\MG(|g|\^{1/2}R+|\hat g|\^{1/2}\hat R), \eeqno{acgmondir}
where $I\^B(\tM\rar\bM)$ is the BIMOND action (\ref{muba}), with $\tM$ replaced by $\bM$.
This equivalent form of the action is useful in separating the effect of MOND, encapsulated in $\tM$ -- attained when $\MG=0$ -- and those of ``$G$ variability,'' encapsulated in $\MG$.
It thus allows us, when considering the resulting field equations, to build on the BIMOND field equations described in Ref. \cite{milgrom22},
and consider only the extra terms due to the added terms in the action (\ref{acgmondir}).
\par
The above action defines a higher-order theory, since $R$ and $\hat R$ depend on second derivatives of the respective metrics.
General relativity is not a higher-order theory, because the dependence on second derivatives in $R$ is such that
\beq |g|\^{1/2} R=|g|\^{1/2}\R+ q\^\n\der{\n},\eeqno{arar}
where
\beq q\^\n=|g|\^{1/2}(\Gmn\C\ten{\l}{\m\l}-\Gma\C\ten{\n}{\m\a}),   \eeqno{quqa}
 and
\beq \R= \Gmn\R_{\m\n}, ~~~~~~~\R_{\m\n}\equiv (\C\ten{\c}{\m\n}\C\ten{\l}{\l\c}-\C\ten{\c}{\m\l}\C\ten{\l}{\n\c}) \eeqno{nirta}
depends only on first derivatives of the metric.\footnote{The definition of $\R_{\m\n}$ here differs by a factor 2 from that in Ref. \cite{milgrom19a}.}
So the higher-derivative part in $R$ is immaterial, as it enters the Lagrangian density as a divergence.
However, in our case, it enters as $\MG q\^\m\der{\m}$ which is genuinely higher order.

\subsection{Field equation   \label{gfield}}
 Varying the integral in Eq. (\ref{acgmondir}) with respect to $\gmn$ and $\hgmn$ we have
\beq -\d_{g}\int d^4x ~|g|\^{1/2}\MG R= \int d^4x~|g|\^{1/2}\d\gmn\left[\MG G^{\m\n}+{\MG}_{,\k}\left(g\ten{\a\k}{}\frac{\d\C\ten{\l}{\a\l}}{\d\gmn}-\Gab\frac{\d\C\ten{\k}{\a\b}}{\d\gmn}\right)
-R\frac{\d\MG}{\d\gmn}\right],\eeqno{dedatur}
and
\beq -\d_{\hat g}\int d^4x ~|\hat g|\^{1/2}\MG \hat R= \int d^4x~|\hat g|\^{1/2}\d\hgmn\left[\MG \hat G^{\m\n}+{\MG}_{,\k}\left(\hat g\ten{\a\k}{}\frac{\d\hat\C\ten{\l}{\a\l}}{\d\hgmn}-\hGab\frac{\d\hat\C\ten{\k}{\a\b}}{\d\hgmn}\right)
-\hat R\frac{\d\MG}{\d\hgmn}\right],\eeqno{dedamak}
where I used the fact that
\beq \d(|g|\^{1/2} R)=|g|\^{1/2} G^{\m\n}\d\gmn+V^\n_{,\n}, ~~~~~~~~~~~  V^\n\equiv |g|\^{1/2}(\Gmn\d\C\ten{\l}{\m\l}-\Gab\d\C\ten{\n}{\a\b}),  \eeqno{shurta}
 and similarly for the twin quantities.
\par
The expressions in square brackets are added to the left-hand side of the BIMOND field equations  (\ref{melte}) and (\ref{melteha}), respectively, where, in addition, $\T\^{\m\n}$ and $\hat \T\^{\m\n}$ in these equations are derived from $\bM$, not from $\tM$.
\par
The first terms, which contributes $\MG G^{\m\n}$ and $\MG \hat G^{\m\n}$, contain second derivatives, but the other terms can be seen to contain third derivatives, since in the Lagrangian density $\MG$ contains first derivatives, while $R$ and $\hat R$ have terms that are only linear in second-derivatives.
\par
While this is a higher-derivative theory -- such as would generically (but  not always) lead to the appearance of Ostrogradsky instabilities \cite{woodard15} -- note that it is a degenerate one, in the sense that is relevant for such theories, since the second derivatives appear linearly in the Lagrangian density. Such degenerate theories are not necessarily beset by Ostrogradsky instabilities. But this needs to be checked specifically the present case for the variety of possible theories in the class. Degenerate, higher-derivative theories, with a function of a scalar field and its first derivative appearing as a prefactor of the Ricci scalar (instead of  $1+\MG$ here) have been discussed, e.g., in Refs. \cite{benachour16,deffayet20}.
\par
Indeed, much remains to be investigated regarding the structure of this theory to assess its merits beyond heuristics.
\par
Another issue to note is the following: Denote, generically, by $\psi$ the various NR potentials; i.e., the small departures of the metric elements from the Minkowski metric in the NR limit.
Then, in the gravitational Lagrangian (\ref{acgmondir}), the dominant term in powers of $\psi$ (in units of $c^2$) up to immaterial derivatives, is of the form
\beq \grad\MG\cdot\grad\psi, \eeqno{narure}
compared with the next-order terms of the form
\beq  \{1+\MG[(\grad\psi)^2/\azs]\}(\grad\psi)^2  ~~~{\rm and}~~~\MG\Delta(\psi^2).  \eeqno{muop}
If indeed it would dominate, such a term would greatly depart from the NR limit of BIMOND itself. However, for high-acceleration systems, the MOND tenets require that $\MG\lll 1 $, and it is easy to make this term strongly subdominant. But, in order to preserve the performance of BIMOND also in low-acceleration systems, such as galaxies, we should also require that $\MG$ vanishes rapidly in the NR limit for all accelerations.
This can be done, as I show in Sec. \ref{examples}, below.
\par
I shall not further expand expressions (\ref{dedatur}) and (\ref{dedamak}), the resulting, cumbersome expressions can be straightforwardly written using formulas in Ref. \cite{milgrom22} (Sec. IIIA there), replacing there $\tM$ by $\MG$ -- the variable on which these functions depend coming from the same stock.
\par
In the examples I give below, in Sec. \ref{constraints}, I consider interesting choices of the form of $\MG$, and relevant circumstances, for which $\MG=\MG^0$ can be taken as a constant. In such cases, we clearly get BIMOND phenomenology, with $G_e=G/(1+\MG^0)$ replacing $G$.

\section{Einstein-Palatini formulation  \label{EP}}
The Einstein-Palatini formulation of GR starts from the Einstein-Hilbert action $\propto \int\gh\Gmn R_{\m\n}$, where the curvature is derived from a connection $\sC\ten{\l}{\m\n}$, symmetric in $\m\n$. In the standard, ``metric,'' formulation of GR, the connection is taken as the Levi-Civita connection, $\C\ten{\l}{\m\n}$, derived from the metric -- thus making the metric the only gravitational DoF. In the Einstein-Palatini formulation of GR, $\sC\ten{\l}{\m\n}$, as it appears in the expression for the curvature scalar, is taken as an independent gravitational DoF.\footnote{A connection appears also in the matter action, via covariant derivatives; there, it is standardly taken as the Levi-Civita, metric connection.} Then, the field equations gotten by varying the action over $\sC\ten{\l}{\m\n}$ tell us that for solutions of the theory we do have $\sC\ten{\l}{\m\n}=\C\ten{\l}{\m\n}$. Then, variation over the metric gives the standard Einstein equations for the metric, resulting in a theory equivalent to GR.
\par
However, when adopting the Einstein-Palatini formulations to modifications of GR [such as to $f(R)$ theories (e.g., \cite{defelice10})], one gets a different theory than is gotten by the standard adoption of the connection as the Levi-Civita one.
\par
We thus consider, in the present context, Einstein-Palatini formulations based on the action (\ref{acgmondir}). A possible advantage of such a (different) theory is that only first derivatives of the DoFs now appear in the Lagrangian density, possibly avoiding the difficulties that may beset the higher-order, metric formulation of Sec. \ref{gfield}.
\par
In our gravitational action, connections appear in the construction of $R$ and $\hat R$, and in the arguments of $\tM$ and $\MG$.  We then have several options for formulating a theory in this vein. The option I shall further follow here involves taking the connections in the arguments of $\tM$ and $\MG$ to still be understood as the Levi-Civita expressions in terms of the metrics (they involve only first derivatives of the metrics), but view the Ricci tensors, $R_{\m\n}$ and $\hat R_{\m\n}$ as constructed from independent-DoF connections $\sC\ten{\l}{\m\n}$ and $\shC\ten{\l}{\m\n}$.
As in the treatment of GR, here too, I take the connections appearing in the matter actions to be the Levi-Civita ones, so matter is still coupled minimally to the metrics. This ensures that the theory tends to GR in the high-acceleration limit, where $\MG\rar 0$ and $\bM,~\tM\rar\tM(\infty)=\O(1)$.
\par
Thus, $\sC\ten{\a}{\m\n}$ and $\shC\ten{\a}{\m\n}$ appear only in the second term in the action (\ref{acgmondir}).
Defining $\sg\_{\m\n}=(1+\MG)\gmn$, and $\hat\sg\_{\m\n}=(1+\MG)\hgmn$, the Einstein-Hilbert part of the gravitational action can be written as
\beq-\frac{c^4}{\spg}\int[\sg^{1/2}\sg\^{\m\n} R_{\m\n}(\sC)+ {\hat\sg}^{1/2}{\hat\sg}\^{\m\n}\hat R_{\m\n}(\shC)].  \eeqno{gamu}

\subsection{Field equation in the Einstein-Palatini formulation  \label{EPfield}}
Equation (\ref{gamu}) tells us that, as in the case of the Einstein-Palatini formulation of GR, varying over $\sC\ten{\a}{\m\n}$ and $\shC\ten{\a}{\m\n}$, their field equations  establish them as the Levi-Civita connections of
$\sg\_{\m\n}$ and $\hat\sg\_{\m\n}$ respectively. Namely,
\beq  \sC\ten{\l}{\m\n}=\oot\sg\^{\l\a}(\sg\_{\m\a,\n}+\sg\_{\n\a,\m}-\sg\_{\m\n,\a})=\C\ten{\l}{\m\n}+\frac{1}{2(1+\MG)}[\d\ten{\l}{\m}{\MG}_{,\n}+\d\ten{\l}{\n}{\MG}_{,\m}   -\gmn g\^{\l\a}{\MG}_{,\a}]\equiv \C\ten{\l}{\m\n}+\G\ten{\l}{\m\n},     \eeqno{natari}
\beq  \shC\ten{\l}{\m\n}=\oot\hat\sg\^{\l\a}(\hat\sg\_{\m\a,\n}+\hat\sg\_{\n\a,\m}-\hat\sg\_{\m\n,\a})= \hC\ten{\l}{\m\n}+\frac{1}{2(1+\MG)}[\d\ten{\l}{\m}{\MG}_{,\n}+\d\ten{\l}{\n}{\MG}_{,\m}   -\hgmn\hat g\^{\l\a}{\MG}_{,\a}]\equiv \hC\ten{\l}{\m\n}+\hat\G\ten{\l}{\m\n},  \eeqno{natamu}
where $\G$ and $\hat\G$ are tensors.
These give $\sC\ten{\a}{\m\n}$ and $\shC\ten{\a}{\m\n}$, each, as an algebraic expression of the two metrics, and of their first and second derivatives.
\par
The relative-accelerations derived from two sets of connections are different, and we have
\beq {^*C}\ten{\l}{\m\n}-C\ten{\l}{\m\n}=-\frac{1}{2(1+\MG)}(\gmn g\^{\l\a}-\hgmn\hat g\^{\l\a}){\MG}_{,\a}=-\frac{1}{2(1+\MG)}(\sg\_{\m\n} \sg\^{\l\a}-\hat\sg\_{\m\n}\hat \sg\^{\l\a}){\MG}_{,\a}.\eeqno{liuty}
\par
As in Sec. \ref{gfield},  I shall not detail the full field equations. Their construction again follows closely the procedure described in In Ref. \cite{milgrom22} (Sec. III A).
\par
Expression (\ref{gamu}) can be viewed as the (uncoupled) Einstein-Hilbert action for the metrics $\sg\_{\m\n}$ and $\hat\sg\_{\m\n}$.
Writing it in this form is the first step in performing the Weyl transformation to the Einstein frame. To complete the transformation we would have to express the metrics and connections appearing in the matter actions and in the BIMOND metric-interaction in terms of $\sg\_{\m\n}$, $\hat\sg\_{\m\n}$, and the starred connections.
In principle, such a transformation may lead to a form of the action that is more amenable to scrutiny of the health of the theory.
 However, it is readily seen that completing the Weyl transformation would be highly involved, and I shall not (and cannot) attempt it here.
 Some of the complications involved concern the fact that this would lead to matter actions in which matter in each sector would appear to couple to both metrics. We also see, e.g., from Eqs. (\ref{natari}) and (\ref{natamu}), that higher derivatives of the metric will enter, via the derivatives of $\MG$. Whether such a transformation would lead to a more useful form is also questionable; for example, its usefulness may be restricted by possible nonanalyticity features of $\MG$ (see Sec. \ref{examples}). All this is also true for the metric formulation of the theory.

\section{Observational constraints on $\MG$ and some examples\label{constraints}}
The full field equations in either formulation of this class of theories are quite complicated.
To boot, there is a large variety of possible theories within the class, according to the choice of scalar variables in both $\tM$ and $\MG$, and the dependence of these functions on them. I shall consider in some more detail only the most simple, but potentially interesting and relevant circumstances where $\MG$ can be taken as nearly constant, $\MG=\MG^0$, to a good enough approximation. The two formulations then coincide, since, in both, the connections that go into the curvature tensors are the Levi-Civita connections of the metrics [see Eqs. (\ref{natari}-\ref{natamu})].
The two formulations then give BIMOND -- still with its variety of choices of $\tM$ -- with an effective $G_e\approx G/(1+\MG^0)$.
\par
Before offering specific examples, I discuss what we require from such theories to achieve in the subcosmological realm, and the resulting constraints on the behavior of $\MG$.

\subsection{Nonrelativistic limit   \label{NRL}}
While other options might be interesting to explore, we take the safe assumption, or requirement, that $\MG$ vanishes rapidly in the NR limit, for both low- and high-acceleration systems.
As already mentioned, this would leave pure BIMOND as governing the dynamics in such systems, retaining its successful performance in galactic phenomenology, including gravitational lensing. We would like to achieve such a behavior of $\MG$ by making it depend in an appropriate way on (``good'') scalar variables of the first and second type.
\par
For this we need to know how these variables behave in the low- and high-acceleration limits. In the post-Newtonian expansion, if we write the MOND potential that governs the dynamics as $\fs=\eps c^2$, then $\eps$ is first order in $(V/c)^2$, where $V$ is the typical virial speed in the system (and I am reinstating $c$ in this section for clarity). For a NR system of mass $M$, and characteristic size $\ell$, $V^2\sim MG/\ell$ in the high-acceleration regime, and $V^2\sim (MG\az)^{1/2}$ in the MOND regime. The $\hsij$ potentials vanish to this order, but, supposedly, they do pick up a post-Newtonian contributions of order $\hsij/c^2=O(\eps^2)$ (or higher, this needs to be checked). This is sufficient to ensure ``correct lensing'' by NR systems\footnote{Since subcosmological, relativistic systems are, perforce, high-acceleration, lensing by them is according to GR, which is the high-acceleration limit of our theories -- see below.} \cite{milgrom22}, since $\grad\hs\ll\grad\fs$ (I use $\grad\hs$ as a generic designation of derivatives of the $\hsij$).
\par
Variables of the first type that are not of the second type, namely those that do contain $\gfss$ terms [$u\not =0$ in Eq. (\ref{bituv})]
are of order\footnote{While, $\fs/c^2$ itself is small, i.e., of order $\eps$, it appears in the arguments of $\tM$ and $\MG$, as $\lM^2\gfss/c^4=\gfss/\az^2$, which is not small in this sense.} $\gfss/\azs$. This is of order $\sim (MG/\ell^2\az)^2\gg 1$ in high-acceleration systems, and of order $\sim MG/\ell^2\az\ll 1$ in low-acceleration ones.
\par
Variables of the second type may have mixed $\hs-\fs$ terms, which contribute $\grad\hs\grad\fs/\azs$ -- which are of at least one order up in $\eps$ -- or pure-$\hs$ terms, which contribute $(\grad\hs/\az)^2$, which are at least two orders up.
However, such second-type variables can be much larger than unity for high-acceleration systems. For example, near the surface of the sun, $\eps\approx 2\times 10^{-6}$, but $MG/\ell^2\az\approx 2\times 10^{12}$, which overwhelms it, yielding $\grad\hs/\az\sim \eps MG/\ell^2\az\gg 1$.
\par
These facts are used in Sec. \ref{examples} to demonstrate that, with appropriate forms of $\MG$, the various observational constraints can be easily satisfied.

\subsection{High-acceleration limit}
There are many and varied, tight constraints on possible inconstancy of $G$.
Most of these constraints come from measurements of $G$ as it enters gravitational dynamics of various subcosmological systems (constraints from cosmology will be considered separately below). These are reviewed and detailed in Ref. \cite{uzan25} (chapter 6).
\par
Examples of such constraints are limits on $\dot G/G$ from lunar and planetary ranging, or from various aspects of pulsar timing. Others constrain the value of $G$ as it entered phenomena related to stellar structure and evolution, such as ages of stars in globular clusters, white-dwarf cooling, and supernova explosions.
\par
In the context of MOND, it is notable that all these subcosmological systems and phenomena -- including other potential ones, such as the emission
of gravitational waves by merging compact objects -- are characterized by accelerations that are many orders of magnitude larger than $\az$.
\par
With the MOND constant at our disposal, we can construct theories that conform to all these constraints, by taking forms of $\MG$ that vanish rapidly in the high-acceleration limit. This would also be in line with the basic tenets of MOND, which dictate return to standard dynamics in this limit, and which already require that the BIMOND interaction, $\tM$, tend to a constant in this limit.
\par
This requirement applies to NR systems -- such as the inner solar system, and noncompact stars -- as well as to relativistic ones -- such as black holes, pulsars, and merging compact objects.
To achieve it, $\MG$ has to depend also on variables that do not vanish in the NR limit -- variables that are of the first type, but not of the second. We can then make $\MG$ vanish rapidly as these variables tend to infinity, independently of the values of the variables of the second type on which $\MG$ depends.

\subsection{Low-acceleration limit}
Possible inconstancy of $G$ is much less constrained in the low-acceleration regime $g\lesssim\az$, accessible, at present, only in galactic systems. After all, it is just such phenomena where standard dynamics strongly fails without invoking dark matter. This leaves the door open for exploring possible appearance of nonvanishing $\MG$ in the low-acceleration regime. But, as indicated above, this is not called for at present, since the NR limit of BIMOND itself appears to account well for galactic dynamics. I shall thus opt here, for the sake of concreteness, to describe examples where $\MG$ vanishes also in the low-acceleration regime. To this end, I make use of the fact that scalars that do not have $\gfss$ terms in their NR limit vanish in this limit to lowest order in $\eps$. In particular, I shall employ variables, $X$, that involve scalars of the second type, and have only $(\grad\hs/\az)^2$ appearing in their NR limit, and which are of order $\eps^2$ in the post-Newtonian expansion.
We see from Eq. (\ref{bituv}), with $u=0$, that such variables form a two-parameter family. Their NR limit is $\bar X=S_{q,0,v}/\azs$, with
\beq \bar S_{q,0,v}=\frac{q}{2}(\hs\_{ij,k}\hs\_{ik,j}-\hs\_{ik,k}\hs\_{ij,j})+v(\hs\_{ik,k}\hs\_{ij,j}+\frac{1}{4}\hs\der{i}\hs\der{i}-\hs\der{i}\hs\_{ik,k}).
\eeqno{biyup}

\subsection{Example  \label{examples}}
To summarize, I want to demonstrate that $\MG$ can be chosen so that the following requirements are satisfied: (1) $\MG$ vanishes rapidly for all high-acceleration systems, relativistic and NR, to comply with the many constraints from subcosmological systems on anomalous $G$ values. (2) $\MG$ also vanishes rapidly in the NR limit even for systems that are not high-acceleration, so that pure-BIMOND phenomenology is predicted for galactic systems. (3) $\MG$ can take finite values in later-time cosmology, while it vanishes for early-time cosmology.
In this section I discuss the fulfilment of requirements (1) and (2) for subcosmological systems. Cosmology will be discussed in Sec. \ref{cosmology}.
\par
In the examples, I use variables of type $X$ described above, and variables, $Y$, built from scalars of the first type, but not of the second; namely, those whose NR limit does contain a $\gfss$ term [$u\not = 0$ in Eq. (\ref{bituv})].
\par
As we saw in Sec. \ref{NRL}, for relativistic, subcosmological systems, of mass $M$, and characteristic size $\ell$, both the $X$ and $Y$ variables are of order $(MG/\ell^2\az)^2\ggg 1$ (because they are perforce of high accelerations).\footnote{For highly relativistic systems, such as the vicinity or a black-hole horizon, $X,Y\sim \ell\_U/\ell$, where $\ell\_U$ is the Hubble distance.} For NR, subcosmological systems, $Y=O(g^2/\azs)$, where $g=MG/\ell^2$ for high-acceleration systems, and $g=(MG\az/\ell^2)^{1/2}$ for low-acceleration ones. And, $X=O(Y\eps^2)$. In the low-acceleration case ($g\not\gg\az$), we have $X\lesssim\eps^2\lll 1$.
To ensure that requirement (2) is satisfied we need $\MG$ to vanish rapidly enough in the low-acceleration case,  as $X\lesssim\eps^2 \rar 0$.

\par
To see how fast, consider the NR limit of the Einstein-Hilbert Lagrangian.
We saw in Eqs. (\ref{narure})-(\ref{muop}) that the post-Newtonically dominant term is
$\grad\MG\cdot\grad\eps$, which we want to be negligible compared with the Poissonian term $(\grad\eps)^2$.
Since $X\lesssim \eps^2$. If $\MG$ vanishes for $X\rar 0$ much faster than $X^{1/2}$ the unwanted $\grad\MG\cdot\grad\eps$ term becomes strongly subdominant in the NR limit.
\par
Thus, construct $\MG(X_1,X_2,...,Y_1,Y_2,...)$, such that it has the following limiting behaviors: (a) It vanishes rapidly when $Y\gg 1$, for all values of $X$ [requirement (1) above]. Then, GR with a cosmological constant is restored for high-acceleration systems (since in this limit we also have $\tM$ going to a constant). (b) It vanishes fast for $X\rar 0$, for all $Y$ values, in particular, even for $Y\not \gg 1$.
\par
For example, using only one variable of each type, take
\beq \MG=-D\z(X)\l(Y),  \eeqno{muteda}
with $\z$ varying (monotonically) from 0 to 1, as $|X|$ varies from 0 to $\infty$, $\l$ varying (monotonically) from 1 and 0, as $|Y|$ varies from 0 to $\infty$, and $D<1$ (to ensure that $G_e=G/(1+\MG)>0$).
One of many possible choices is \beq  \z(X)=\exp(-\verb"a"/X^2)~~~~{\rm and}~~~~\l(Y)=\exp(-\verb"b"Y^2).   \eeqno{molaret}
Then, all the above requirements are satisfied, since $\MG$ vanishes rapidly for $Y\gg \verb"b"^{-1/2}$ for all $X$, and for $X\ll \verb"a"^{1/2}$ for all $Y$.
The values of the scalars can have both signs, depending on the circumstances; e.g., the same scalar can have different signs in cosmology, where time dependence is important, and in local static systems, where time dependence is negligible. I take these functions to depend on $X^2$ and $Y^2$, only because I do not want to worry about the signs of the variables, at the present level of demonstration. In more realistic examples, it may be useful to also employ the difference in sign.
\par
With such a choice, $\MG$ makes itself felt only for systems that are relativistic and that are not characterized by accelerations much larger than $\az$. This leaves us only with cosmology as a self gravitating system that may be affected by $G$ variability. In the above example, $\MG$ can be between 0 and $-D$; so, if $D\not \ll 1$ the corresponding $G_e>G$.

\par
The expression for $\MG$ in the example above is nonanalytic in $Y$ in the high-acceleration limit -- it cannot be expanded in powers of $\az$ near the GR limit $\az=0$.
It is also nonanalytic in the variable $X$ near the decoupling limit $X=0$ (also a GR limit with a different CC).
This nonanalyticity can stand in the way of applying some standard techniques to assess the health of the theory. But this, in itself, does not argue against such a behavior. In quantum theory, for example, we encounter expression -- such as the barrier-penetration probability, or the path weight in the path-integral approach -- that have similar nonanalytic dependence on $\hbar$ at the classical limit $\hbar=0$.
(MOND is also nonanalytic in the deep-MOND limit due to the appearance of fractional powers in the Lagrangian -- forced by phenomenology.)
\par
In fact, I choose this example behavior intentionally for two reasons. The first is that phenomenological constraints -- in particular solar-system constraints -- require that the GR limit is approached very fast in the high-acceleration limit ($\az\rar 0$), in $\MG$ as in $\tM$. Also, I wanted an example of $\MG$ that vanishes fast for small $X$, so as to be compatible with BBN limits, on one hand, and, on the other, to grow rapidly to its asymptotic constant value, so as to account for faux-DM effects in the expansion history. All this can be achieved perhaps with analytic functions, which, however, I do not see as a pressing desideratum. The second, and stronger, reason is that there exist general theorems to the effect that bimetric theories, generically, possess unwanted ghost instabilities (see, e.g., Refs. \cite{boulanger01,dR11,dambrosio20}).
However, such theorems build on expansions in the metric coupling near the decoupling limit; so, opting for nonanalytic behavior may escape the sentence of such theorems, even though, in itself, it does not guarantee the absence of ghosts. (See also the comments on the possible presence of ghosts in VGMOND and in BIMOND generally, in Sec. \ref{introduction}.)

\section{Cosmology \label{cosmology}}
Our present picture of cosmology within GR has not emerged as a prediction of the theory -- general relativity can, of course, accommodate many other consistent cosmological models. This picture has been built layer upon layer with the benefit of major inputs beyond the adoption of GR: Inputs from physics -- as our view of its laws have evolved; inputs from observations -- as these have unfolded -- e.g., on the realization that galaxies are ``island universes,'' on the expansion of the universe, etc.
The material contents of the universe, with their properties and interactions, are all put in by hand. This includes components -- dark matter and dark energy -- that we do not even know exist from independent evidence, and whose presence, amounts, distributions, exact nature, and interactions are all free for one to invoke as long as they help square GR with the observations. Symmetries -- the cosmological principle -- are indicated by observations, but then assumed generally. Initial conditions, and whether, and how they follow from inflation, in any of its many versions, are still moot.
\par
The situation in BIMOND and \VGM~ is even more involved. We have no information on all the above aspects of the twin sector: For example, on the possible existence of matter in the twin sector, on its amounts, and on its physical properties. Likewise, we are ignorant of the initial conditions of this sector. To these is added the fact that BIMOND and \VGM~ are not a single theory, but classes of theories, with different possible choices of the dependence of $\tM$ and $\MG$ on the scalar variables.
\par
Similarly to how we have painstakingly proceeded with GR cosmology, the approach one can take here is to see if there is some version of the theory, and some reasonable set of inputs concerning the material contents, and the cosmological initial conditions in the two sectors, that can account for the observed cosmology without introducing dark matter and ``material'' dark  energy. This would take much work to explore within BIMOND/\VGM.
\par
In this spirit, references \cite{milgrom10a,clifton10}, studied cosmology in BIMOND; however, they considered only BIMOND itself, with its original form, which employed only one specific scalar variable -- the contraction of $\Up\_{\m\n}\equiv C\ten{\c}{\m\l}C\ten{\l}{\n\c}-C\ten{\l}{\m\n}C\ten{\a}{\l\a}$. In the meanwhile, we have realized \cite{milgrom22} that BIMOND theories with a much larger variety of scalar variables are consistent with gravitational lensing. And, here we also have to consider the variety of choices of $\MG$, which are only weakly constrained by galactic dynamics. This gives us a much larger space of theories and cosmological inputs to consider.
\par
The main constraint on variability of $G$ from cosmology that is relevant to our discussion here comes from big bang nucleosynthesis (BBN), and emerges essentially from the observed abundances of the light elements, which constrain the expansion rate at the relevant time. Reference \cite{uzan25} discusses such constraints, with the various caveats that underlie them, and concludes: `` ... assuming the number of neutrinos to be three, leads to the conclusion
that $G$ has not varied from more than 20\% since nucleosynthesis. But, allowing for a change both in $G$ and $N\n$ allows one for a wider range of variation.
\par
Other constraints may come from observations and interpretation of the CMB. But, according to Ref. \cite{uzan25}: ``Cosmological observations are more difficult to use in order to set constraints
on the time variation of $G$. In particular, they require to have some ideas about
the whole history of $G$ as a function of time but also, as the variation of $G$
reflects an extension of general relativity, it requires to modify all equations
describing the evolutions of the universe and of the large-scale structure in
a consistent way.'' And more from Ref. \cite{uzan25}: ``In full generality, the variation of $G$ on the CMB temperature anisotropies depends on many factors: (1) modification of the background equations and the evolution of the universe, (2) modification of the perturbation equations, (3) whether the scalar field inducing the time variation of
$G$ is negligible or not compared to the other matter components, (4) on the time profile of $G$ that
has to be determine to be consistent with the other equations of evolution. This explains why it
is very difficult to state a definitive constraint.''
\par
All this clearly applies to \VGM.
\par
In light of the BBN constraint, one would require from an acceptable solution of the theory to yield $\MG\ll 1$ during BBN,  but $\MG\sim -1$ later on, when ``dark matter'' effects are needed to account for the expansion history; in particular, during matter dominance\footnote{Without dark matter, matter dominance occurs later than in the standard picture.}.
\par
The two epochs differ in several regards that could lead to such a variation in the value of $\MG$: (a) The former epoch is radiation dominated, while the latter is essentially matter dominated. (b) There may have been differences between the metrics in the two sectors, as the Universe emerges from a period of inflations, e.g., in the form of scalar or tensor (graviton) waves \cite{woodard18,woodard24}. These could
have contributed to the \VGM~ scalar variables, with changes between the two epochs. (c) Depending on the initial conditions in the two sectors, the former epoch may have been characterized by very near metric symmetry, hence near vanishing of the scalar variables, while later on asymmetries develop due to independent fluctuations on top of the common FLRW metric, leading to finite values of the scalar variables.
\par
For the sake of concreteness, but also because it makes sense on various grounds, I am assuming, when considering BIMOND and \VGM~ cosmology in what follows, that the two sectors are exactly symmetric as regards their underlying physical laws and their matter components, with all their physical properties and interactions.
\par
I also assume symmetric initial conditions in the two sectors. Differences between the two sectors can develop at later times, inasmuch as they result from random processes, which can occur independently in the two sectors. And because such fluctuations in the two sectors do interact (only) gravitationally through the BIMOND interaction terms, they can even greatly affect each other's development, as discussed in detail in Ref. \cite{milgrom10b}.
\par
Another possible emergent difference between the two sectors, which I mention only in passing, and which I shall not further pursue here, is in the number of baryons. We do not know how, and when exactly, the amount of baryons in our sector was fixed. The standard assumption is that the universe started with a zero baryon number, and at some point, when the three Sakharov conditions \cite{sakharov67} were fulfilled, a very small number imbalance between baryons and antibaryons was created, which is what appears today as the remaining baryons after the vast majority of baryons and antibaryons had annihilated.
Depending on the exact mechanism responsible for the slight baryon asymmetry, this small difference could have been different in the two sectors, leading to a difference in the densities of baryons and ``twin baryons.''
\par
As somewhat of an aside, I note that one may object that in such a picture as I just described, the gain in obviating dark matter is annulled by having to postulate the existence of the even-more-speculative ``twin matter''. This is, however a specious objection. The main deficiency of the dark-matter paradigm is not that it invokes the omnipresence of some material component that is not part of known physics. It is that within this paradigm one does not expect -- to say nothing of predict -- the very clear-cut laws and regularities that are obeyed by galactic dynamics.\footnote{Claims that such regularities emerge from simulations of galaxy formation in \LCDM are highly misleading. They are all essentially put in by hand, by setting the many free dials that control these simulations to match as far as possible observed galaxies. These simulations could produce very different galaxies if observations required it.}
In contradistinction, the main gain in MOND is not that it eliminates dark matter; it is the fact that the observed regularities follow, and were predicted, as unavoidable consequences of the theory -- they are the ``MOND laws of galactic dynamics'' \cite{milgrom14a} -- as Kepler's laws follow from Newtonian dynamics. These laws must be obeyed by all galactic system, whatever their complicated formation and evolution have been. {\it This fact is not changed by invoking the existence of twin matter in a {\rm BIMOND} theory.} Twin matter does not play the role of dark matter, and, in fact, need not, and should not, be present around present day galaxies, and has no effect on their observed dynamics \cite{milgrom09b,milgrom10a}.
\par
As a result of the assumed symmetry between the two sectors, they have the same global cosmological evolution, with the space averages of the two metrics remaining the same\footnote{Note that the scalar variables do not average to zero, even if the metric difference itself does.}. With the standard cosmological principle assumed, the common cosmological metric has the
FLRW form.
\par
With this in mind, we can write the two metrics as
\beq \gmn=\rgmn +\hmn,~~~~ \hgmn=\rgmn+\hat h\_{\m\n},  \eeqno{cosmfl}
where $\rgmn$ is of the FLRW form.
\par
We shall, further, consider cosmologies in which $\hmn$ and $\hat h\_{\m\n}$ are small and can be treated to lowest order.
So, for example, to this order $\Gmn=\tilde g\^{\m\n}-\tilde g\^{\m\a}\tilde g\^{\n\b}h\_{\a\b}$.
Defining
\beq  \hs\_{\m\n}=h\_{\m\n}-\hat h\_{\m\n},  \eeqno{maza}
these will appear in the scalar variables to the same low order.
We shall also consider cases where $\hmn$, $\hat h\_{\m\n}, \hs\_{\m\n}$ vary on spatiotemporal scales much smaller than the corresponding cosmological ones; so, for example, we will neglect expressions of the form $\rgmn{\_{,\l}}h\_{\a\b}$ compared with $\rgmn h\_{\a\b,\l}$.
This also means that we can raise and lower indices of $\hmn$, $\hat h\_{\m\n}$, and $\hs\_{\m\n}$ with $\rgmn$  inside derivatives; e.g., $\tilde g\^{\m\n}h\_{\n\l,\a}=(\tilde g\^{\m\n}h\_{\n\l})\_{,\a} \equiv h\ten{\m}{\l,\a}$.
\par
This is another case where Eq. (\ref{meqaz}) can be used, from which one calculates the following expressions for the five independent, basic
scalars:

$$ S_1=\frac{1}{4}{\hs}\^{\m\n,\a}(2{\hs}\_{\m\a,\n}-{\hs}\_{\m\n,\a}),~~~~~~
S_2=\oot {\hs}\^{,\m}({\hs}\^\n\_{\m,\n}-\oot {\hs}\der{\m}), ~~~~~~
 S_3=({{\hs}\^{\m\n}}\_{,\n}-\oot {\hs}\^{,\m})({\hs}\^{\l}\_{\m,\l}-\oot {\hs}\der{\m}),  $$
\beq S_4=\frac{1}{4}{\hs}\^{,\m}{\hs}\_{,\m},~~~~~~
 S_5=\frac{1}{4}{\hs}\^{\m\n,\a}(3{\hs}\_{\m\n,\a}-2{\hs}\_{\m\a,\n}).\eeqno{s5m}

Here, ${\hs}={{\hs}}\ten{\m}{\m}$ is the (4-dimensional) trace of $\hs\_{\m\n}$, and upper indices were raised with $\tilde g\^{\m\n}$.
For such a double geometry, the good scalars of the first kind, which we use in $\tM$ and $\MG$ are of thee form
\beq S_{q,u,v}=\frac{u}{16}(2{\hs}\^{\m\n,\a}{\hs}\_{\m\n,\a}-{\hs}\^{,\m}{\hs}\_{,\m})
+\frac{q}{2}({\hs}\^{\m\n,\a}{\hs}\_{\m\a,\n}-{{\hs}\^{\m\n}}\_{,\n}{\hs}\^{\l}\_{\m,\l})
+v({{\hs}\^{\m\n}}\_{,\n}-\oot {\hs}\^{,\m})({\hs}\^{\l}\_{\m,\l}-\oot {\hs}\der{\m}).  \eeqno{birbur}
The subfamily of these, the $X$ scalars of the second type that we employed in the examples of Sec. \ref{examples} as arguments of $\MG$, have $u=0$.
\par
It may also be of use to note that $\hs\_{\m\n}$ that
 satisfies the harmonic gauge annihilates such $X$ scalars for which, in addition, $q=0$. It was shown in Ref. \cite{milgrom14b} that this holds for any package of plane gravitational waves under the harmonic gauge.
\par
In a scenario such as described above, initially, due to the exact symmetry, $\hs\_{\m\n}=0$, and all the relative-acceleration scalars vanish.
We saw that the observed dynamics of subcosmological systems indicates that $\MG=0$ in this case.
We then have a cosmological solution where the two metrics are equal, and are described by the GR, FLRW geometry with a cosmological constant $\Lambda= -\lM^{-2}\tM_0/2$, where $\tM_0$ is the value of $\tM$ with all its argument set to zero.
\par
At later times the symmetry is broken, as described above, and the scalars become finite, and $\tM$ and $\MG$ become spatially and temporally variable.
\par
On small scales, the interaction term accounts for MOND effect in structure formation, and later in the dynamics of well-formed galactic systems. In the latter, approximately-static systems, it is the derivatives of the interaction with respect to the variable that enter MOND dynamics \cite{milgrom09b,milgrom22}, and define the appropriate MOND ``interpolating function''. A constant contribution to the interaction, $\tM\_{DE}$, has a minor effect on the dynamics in systems small compared with $\lM$, if $\tM\_{DE}=O(1)$.
\par
When referring  to a ``cosmological constant'' within the standard model of cosmology, we mean the constant $\Lambda$ that enters the Einstein equation as
$G^{\m\n}=-8\pi G T^{\m\n}-\Lambda\Gmn$, where $T^{\m\n}$ is the energy-momentum tensor for matter, and $G$ takes the standard value.
If this is not a good description of the real world, possible ambiguity may appear as to what we count as dark energy and what as anomalously-behaving matter.
The distinction between the contribution of matter and cosmological constant is also clear in our \VGM~ case when $\MG$ and $\tM$ are identically constants. Then the field equations for the metrics are
\beq G^{\m\n}=-\frac{8\pi G}{1+\MG} T^{\m\n}+\frac{\tM}{2\lM^2}\Gmn, \eeqno{mareq}
and similarly for $\hgmn$.
Observing such a behavior (e.g., in the expansion history), and interpreting it based on GR, we would say that we have $(1+\MG)^{-1}$ times the actual contribution of matter (attributing the excess to dark matter), and identify $-\tM/2\lM^2$ as a cosmological constant.
\par
But, generally, when  $\MG$ and $\tM$ cannot be treated as constants, the distinction between the extra strength of gravity associated with matter, and the contribution of a dark energy term is not clear cut.
\par
Even within strict BIMOND ($\MG\equiv 0$), because the metrics appear in the scalar variables, the interaction term, $\tM$, gives rise in the field equations to terms that are proportional to $\tM\Gmn$ and $\tM\hGmn$, and so have an effective ``equation of state $p=-\r$, with $\r$ somewhat space-time dependent. But it also gives rise to terms that are not proportional to $\Gmn$ or $\hGmn$ (see Refs. \cite{milgrom09b,milgrom22}). So, if it is legitimate to take spatial averages of all quantities in the field equation when describing the expansion history we would get an equation of state that describes dark energy.
\par
We have no {\it a priori} idea what sort of differences between the metrics may develop, and what values the scalar variables may then attain. Such differences depend on unknown initial conditions, and their subsequent development has to be determined consistently from the theory at hand -- the version of \VGM~ under consideration.
\par
But, just as a heuristic example, to demonstrate how the different scaler variables may have different behaviors and contributions in the NR limit and in cosmology, I end this section by considering a specific ansatz for $\hs\_{\m\n}$, which was used in Ref. \cite{milgrom22} for another purpose.
In this example, the FLRW metric has zero spatial curvature:
\beq \rgmn={\rm diag}(-1,a\^2,a\^2,a\^2), \eeqno{cosmara}
with $a(t)$ the cosmological scale factor, and invoke, in analogy with the form of the metrics in the NR limit, an ansatz where
\beq \hmn= -2\t~ {\rm diag}(1,a\^2,a\^2,a\^2),~~~~ \hat h\_{\m\n}= -2\hat\t~ {\rm diag}(1,a\^2,a\^2,a\^2).  \eeqno{cosmfer}
As indicated above, $\t$ and $\hat \t$ are assumed much smaller than 1 (in units of $c^2$), but their derivatives are not assumed small (in units of $\az$). Their spatial wavelengths are assumed much smaller than the cosmological horizon, and frequencies much larger than the (local) expansion rate.
\par
With these strong inequalities (which allow us, for example, to neglect $\t\tilde g\_{\m\n,\l}$ relative to $\t\_{,\l}\tilde g\_{\m\n}$, etc.) we have for the five basic scalars (contracted with the reference metric)
\beq S_1=-2\rGmn\vrf\der{\m}\vrf\der{\n};~~~~~S_2=8\vrfzs;~~~~~ S_3=-16\vrfzs;~~~~~ S_4=4\rGmn\vrf\der{\m}\vrf\der{\n};~~~~~S_5=10\rGmn\vrf\der{\m}\vrf\der{\n},  \eeqno{lituret}
where, $\varphi\equiv\t-\hat\t$, $\rGmn\vrf\der{\m}\vrf\der{\n}= -\vrfzs +\gvrfs$, and $\nabla\vrf$ is the space gradient of $\vrf$ with respect to the proper distances $dl=adx$.
Then, we find for the good scalars
\beq   S_{q,u,v}=-16v\vrfzs +u\rGmn\vrf\der{\m}\vrf\der{\n}.  \eeqno{numirafet}
We see in this example that scalars of the second type, characterized by $u=0$, and which disappear from the NR limit, can still be important in cosmology due to the time dependence of the metric difference. On the other hand, scalars of the first type ($u\not =0$), but with $v=0$, which
fully contribute in NR systems, may vanish in this example, if $\varphi\_{,\m}$ is lightlike.
So, we may envisage, for example, a \VGM~ version where the interaction depends on two scalars $X$ and $Y$, with $Y$ having $u\not=0$, and $X$ having $u=0$, with the dependence of $\tM$ and $\MG$ on them being such that the dependence on $X$ enters cosmology, and that on $Y$ determines the NR limit.

\section{Discussion   \label{discussion}}
I have proposed, and demonstrated with the specific example of BIMOND, that MOND offers a framework for variable-$G$ theories that can naturally comply, without fine tuning, with all the constraints from subcosmological systems on $G$ inconstancy, and yet exhibit variable-$G$ effects in cosmology.
\par
The added, variable-$G$ aspects of such a theory are not needed, and perhaps could stand in the way, in applications to galactic dynamics. I have thus concentrated on versions of the theory that do not predict any such effects in galactic systems.
In such systems we do have good constraints from observations on the type of scalar variables that can appear, and on the dependence of the interaction term on them \cite{milgrom22}.
In contradistinction, in cosmology, where the main motivation for introducing variable $G$ lies, we are hardly constrained, since all variables can contribute and take up different values according to circumstances. I am thus not able to offer specific promising scenarios for cosmology in this framework, beyond laying the grounds for the study of such theories.
\par
Extensions in a similar vein can be constructed for other relativistic MOND theories, such as RMOND/AeST \cite{sz21}, and the Khronon-Tensor theory described in Ref. \cite{blanchet24}, by multiplying the Ricci scalar in their Einstein-Hilbert action by a function of the gradient of the scalar degree of freedom of the theory.
However, these two theories account for the effect on the expansion history that are conventionally attributed to dark matter in their own way, and do not requite variable-$G$ effects for this purpose.
\par
There are important remaining question regarding the proposed class of BIMOND-based \VGM~ theories that require elaboration.
In particular, it is not clear that a consistent cosmology can be found within this framework that is consistent with all observations and that does away with the need for dark matter.
\par
The possible effects on gravitational-wave propagation constitute another issue that needs to be addressed. Such effects depend strongly on the choice of variables, and the dependence of $\MG$ on them.
A possibly relevant finding is that, as was shown in Ref. \cite{milgrom14b}, and alluded to above, below Eq. (\ref{birbur}), there are ``good'' scalar variables that vanish for any packet of plane gravitational waves that satisfy the harmonic gauge.
\par
The possible presence of ghosts, which looms over BIMOND itself, and the added question of Ostrogradsky instabilities in the affine and in the Einstein-Palatini formulations of \VGM~ has to be investigated in detail. We need to check if such unwanted feature exist, and if they do, how deleterious they are, and whether there are ways to exorcise them.

{\bf Acknowledgements:} I thank the reviewer for useful suggestions.

\end{document}